# The NTNU Taiwanese ASR System for Formosa Speech Recognition Challenge 2020

Fu-An Chao[1], Tien-Hong Lo[1], Shi-Yan Weng[1], Shih-Hsuan Chiu[1],
Yao-Ting Sung[2,3], Berlin Chen[1]

[1]Department of Computer Science and Information Engineering,
National Taiwan Normal University, Taipei, Taiwan

[2]Department of Educational Psychology and Counseling,
National Taiwan Normal University, Taipei, Taiwan

[3]Institute for Research Excellence in Learning Sciences,
National Taiwan Normal University, Taipei, Taiwan

{fuann, tienhonglo, 60947007S, shchiu, sungtc, berlin}@ntnu.edu.tw

## Abstract

This paper describes the NTNU ASR system participating in the Formosa Speech Recognition Challenge 2020 (FSR-2020) supported by the Formosa Speech in the Wild project (FSW). FSR-2020 aims at fostering the development of Taiwanese speech recognition. Apart from the issues on tonal and dialectical variations of the Taiwanese language, speech artificially contaminated with different types of real-world noise also has to be dealt with in the final test stage; all of these make FSR-2020 much more challenging than before. To work around the under-resourced issue, the main technical aspects of our ASR system include various deep learning techniques, such as transfer learning, semi-supervised learning, front-end speech enhancement and model ensemble, as well as data cleansing and data augmentation conducted on the training data. With the best configuration, our system obtains 13.1 % syllable error rate (SER) on the final-test set, achieving the first place among all participating systems on Track 3.

**Keywords**: Formosa Speech Recognition Challenge, deep learning, transfer learning, semi-supervised training

# 1. Introduction

Due to the rapid developments of deep learning, deep neural network (DNN) based techniques have enjoyed widespread adoption in the automatic speech recognition (ASR) community. ASR has also found its applications in many different areas, ranging from interactive voice response (IVR) services and personal assistants, for which people can interact with the machine naturally by using their own voices, meeting transcription, speech translation to speech summarization. Nowadays, some top-of-the-line ASR systems can even reach the performance level of professional human annotators in English, a dominant language in the world. However, in real-world scenarios, there exist some languages that are resource-poor or even endangered. For example, although both Mandarin (a.k.a. Pǔtōnghuà and Huáyǔ) and Taiwanese (a.k.a. Taiwanese Hokkien, Hoklo, Taigi, Southern Min and Min-Nan) are Chinese dialects spoken by large populations of people, the latter is underexplored and has far less ASR training data made publicly available than the former, which causes the performance of Taiwanese ASR systems to fall short of expectations. Furthermore, distinct from Mandarin, Taiwanese has a wide variety of pronunciation traits that can be attributed to the influences from disparate languages like Formosan, Dutch, Japanese, and among others [1]. Despite there are still 70% of the population in Taiwan who use Taiwanese to communicate, most of the people, young generations in particular, have only limited vocabulary and cannot speak Taiwanese fluently. Therefore, in addition to the continued promotion of the Taiwanese language and the preservation of its associated culture in the 21st century, how to empower Taiwanese ASR applications in daily life, like voice command control and automatic TV show subtitling and IVR, to name just a few, remains to be of prime importance.

This paper describes the NTNU ASR system participating in the Formosa Speech Recognition challenge 2020 (FSR-2020) supported by the Formosa Speech in the Wild[1] project (FSW). Figure 1 outlines the major components of our system submitted to FSR-2020. The high variety existing in the pronunciation characteristics of Taiwanese and the varying noise-contaminated test conditions of the FSR-2020 datasets make this challenge intrinsically much more difficult. In the setting of the Track 3 competition, the output hypotheses of an ASR system have to be tonal syllable sequences, with a tone index, ranging from 1 to 9, attached to each syllable. Apart from the training data provided by the organizer, all participants were allowed to build their systems with additional data outside the FSR-2020 training dataset.

---

[1] https://sites.google.com/speech.ntut.edu.tw/fsw

Instead of recourse to extra in-domain speech data, we stick to conducting Taiwanese ASR on a resource-scarce assumption. To this end, we explore the joint use of several ASR modeling strategies, including data augmentation, transfer learning, semi-supervised training and model ensemble. In addition, to alleviate the negative effects of ambient noise and reverberation that may mix with the test utterances of Track 3, speech enhancement is also applied to generate augmented data for training the acoustic models. Finally, our system with the best configuration takes the first place among all participating systems on Track 3.

The remainder of this paper is organized as follows: Section 2 sheds light on our main contributions and the strategies that were employed, from front-end processing to back-end acoustic and language modeling. Section 3 presents the details of the experimental setup, results and discussion. Finally, we conclude the paper and envisage future research directions in Section 4.

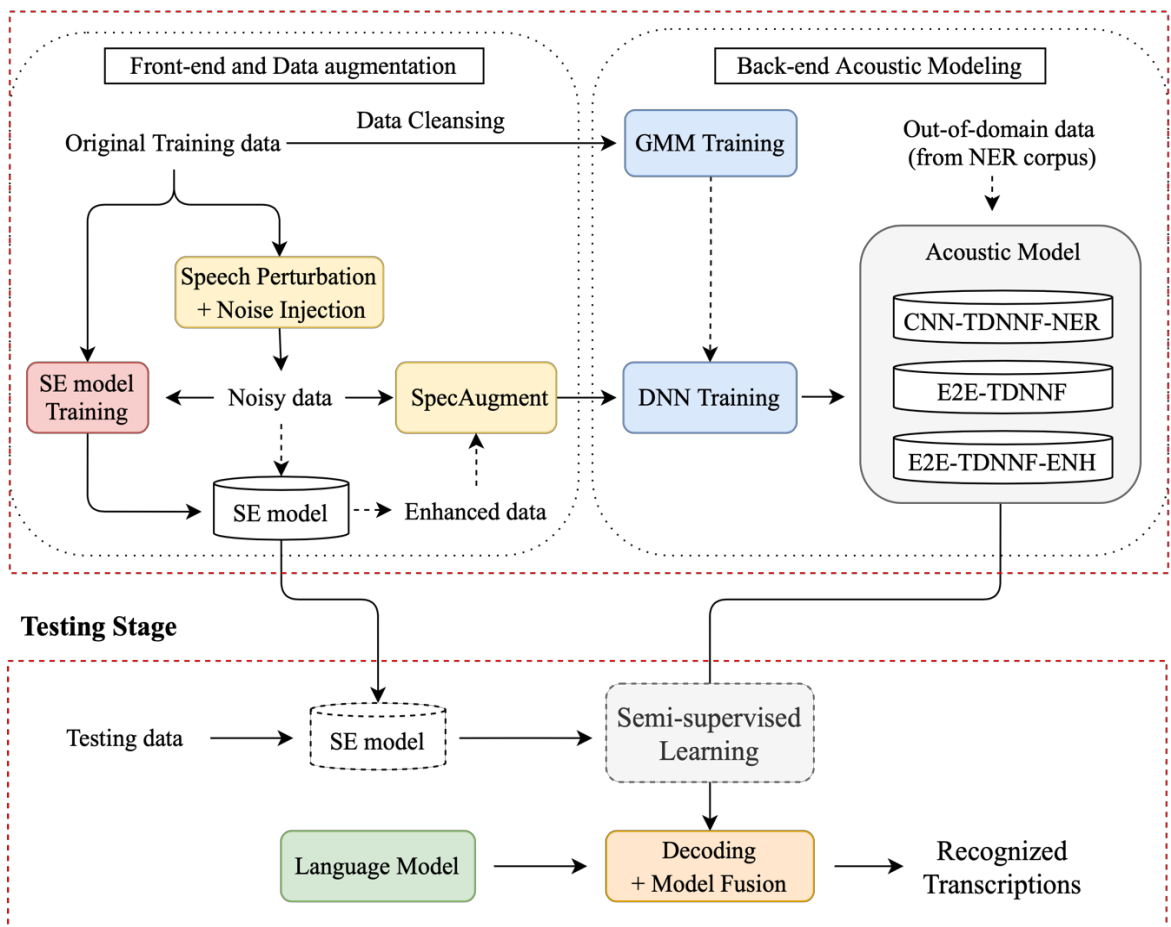

Figure 1: An overview of the NTNU system for FSR-2020.

## 2. Strategies for Building a Taiwanese ASR System

In this section, we present our main strategies for building a Taiwanese ASR system for FSR-2020, which consist of two preprocessing procedures, viz. lexicon augmentation and data cleansing, training data augmentation, front-end speech enhancement, acoustic modeling and language modeling. Each of the aforementioned components will be elaborated on in the following subsections, respectively.

### 2.1 Lexicon augmentation and data cleansing

First of all, through a careful inspection of the training dataset and baseline recipe[2] provided by the FSR-2020 organizer, we noticed there were a few syllable patterns in the given lexicon missing their pronunciations. In order to make the lexicon more complete, we used a greedy approach to performing lexicon augmentation. Specifically, we first enumerated all distinct tonal syllables that appeared in the reference transcripts of the training dataset and then augmented the lexicon with those tonal syllables that were not found in the lexicon. As a side note, we excluded none-Taiwanese lexical patterns that appeared in the transcripts of the training dataset, such as English words and proper nouns (e.g., Google), from consideration in this study.

The acoustic model of our ASR system was built with a hybrid deep neural network and hidden Markov model (DNN-HMM) structure, which employed a DNN in place of Gaussian mixture model (GMM) for modeling the state emission probabilities in a traditional GMM-HMM structure. Hybrid DNN-HMM acoustic models have shown to be significantly superior than the conventional GMM-HMM acoustic models on many ASR tasks. It is arguable that Hybrid DNN-HMM acoustic models still have to resort to GMM-HMM acoustic models to obtain good forced-alignment information for better estimation of their corresponding neural network parameters. Inspired by this practice, the GMM-HMM acoustic model of our best system was trained with the audio segments that were screened out from speech training dataset with high recognition confidence scores generated by an existing hybrid DNN-HMM system. As we shall see later, the empirical ASR results confirm this intuitive data-cleansing therapy.

### 2.2 Data augmentation

The training dataset for the Track 3 of FSR-2020 was provided by the organizer, which consisted of about 50 hours of Taiwanese utterances. This amount of dataset would be insufficient when training a hybrid DNN-HMM acoustic model for Taiwanese ASR. To enrich the speech training data and increase the robustness of our ASR system for the Track 3 competition, we thus set out to leverage various data augmentation strategies based on different label-preserving transformations.

---
[2] https://github.com/t108368084/Taiwanese-Speech-Recognition-Recipe

In addition to utterance-level speed perturbation [3] used in our baseline system, we also adopted other data augmentation methods, including spectrogram augmentation and noise injection. As we shall see in the experimental section, these data augmentation strategies collectively lead to promising results that further push the limits of our ASR system.

2.2.1 Spectrogram augmentation

Apart from the augmentation strategies that operate in the waveform domain [3], feature-based augmentation that is conducted in the feature-space domain (e.g., spectrum or spectrogram) is another active line of research for acoustic modeling in past few years. One of the most celebrated feature-space augmentation methods adopted for acoustic modeling is vocal tract length perturbation (VTLP) [4]. VTLP employs a linear warping transformation along the frequency bins, simulating the effect of altering the vocal tract lengths of speakers that produce the training utterances. More recently, SpecAugment [5] has drawn much attention from the ASR community. With the inspiration from computer vision (CV), SpecAugment treats the spectrogram of an utterance as an image, which first performs warping along the time axis (time-warping) and then masks blocks of consecutive time and frequency bins in different axes (time-frequency masking). The whole operations of SpecAugment jointly lead to considerable word error rate reductions on several benchmark ASR tasks.

In this paper, we made use of the component "spec-augment-layer" of the Kaldi toolkit [16] along with speed perturbation to generating label-preserving, augmented data for training our acoustic model. Note here that "spec-augment-layer" consists of only time and frequency masking operations. This is probably because the time-warping operation is conceptually similar to speed perturbation conducted in the waveform domain, but costs a larger amount of computation and may not lead to substantial improvements [5].

2.2.3 Noise injection

To alleviate the deteriorating effects of time- or frequency-varying noise when testing our system in an unseen environment, we also injected different types of noise into the training speech utterances, which were compiled from a few online-available noise datasets

[18][19][20][21][22]. In doing so, we can not only increase the diversity of training data but also prevent the hybrid DNN-HMM acoustic model from encountering the overfitting problem. Specifically, we randomly selected a signal-to-noise-ratio (SNR) ranging from -5 to 15 dB when contaminating each speed-perturbed utterance with a certain type of noise, totally creating a 6-fold augmentation of the training dataset for estimating the acoustic model. In addition, these corrupted utterances were also used to train our front-end speech enhancement component (see Section 2.3 that follows).

**2.3 Front-end speech enhancement**

In a wide variety of realistic situations, the input to an ASR system might be noise-contaminated speech. As a solution to mitigate the undesirable noise-contamination effects, speech enhancement (SE) is arguably a crucial modeling paradigm to improve noise robustness of acoustic modeling. Particularly, time-domain SE methods have drawn much attention from both the academic and commercial sectors in past few years and have exhibited outstanding noise-reduction performance on many ASR tasks.

As such, we used the fully-convolutional time-domain audio separation network (Conv-TasNet) [6] as a preprocessing component for noise suppression, which was originally proposed for the speech separation task intended to separate an input mixture to individual speech signals. Conv-TasNet has shown superior performance over many frequency-domain approaches. The main architecture of Conv-TasNet is composed of an encoder, decoder and masking network, jointly processing a raw waveform signal in an end-to-end manner. As we shall see later, with the adoption of Conv-TasNet, we can obtain marked improvements in both front-end and back-end ASR evaluations.

**2.4 Acoustic modeling**

The DNN component of our hybrid DNN-HMM acoustic model involves several layers of factorized time-delay neural network (TDNNF) [2], optionally prepended by several layers of convolutional neural network (CNN). Such a pairing of neural networks is denoted by CNN-TDNNF hereafter. TDNNF is viewed as an effective extension to TDNN (time-delay neural network), with the purpose of obtaining better modeling performance and meanwhile reducing the number of parameters by factorizing the weight matrix of each TDNN layer into a product of two low-rank matrices. As a side note, it is worth mentioning that our hybrid DNN-HMM acoustic model can also be estimated with the so-called flat-start, end-to-end training setup suggested by [8] (denoted by E2E-TDNNF hereafter). This setup facilitates the training of a hybrid DNN-HMM acoustic model without resort to any previously trained acoustic models and forced-alignment information, while the estimated model can still work pretty well especially in noisy test scenarios.

In addition, the objective function for training the acoustic model is lattice-free maximum mutual information (LF-MMI) [7]

$$\mathcal{F}_{\text{LFMMI}} = \sum_{i=1}^{N} \log \frac{P(\mathbf{O}_i|L_i)^k P(L_i)}{\sum_L P(\mathbf{O}_i|L)^k P(L)} \tag{1}$$

where $\mathbf{O}_i$ and $L_i$ are the acoustic feature vector sequence and the corresponding phone sequence of the $i$-th training utterance, $k$ is a weighting factor, and $P(L_i)$ is the phone $N$-gram language model probability.

### 2.5 Language modeling

For language modeling, we first adopted the SRILM toolkit [9] to train $N$-gram based language models, for which both the *Good-Turing* [10] and *Kneser-Ney* [11] $N$-gram smoothing methods were considered. In our experiments, we observed that using a four-gram language model yielded considerable improvements than a tri-gram language model in terms of both the perplexity and ASR error reductions. Also worth mentioning is that both these language models were trained solely on the text corpus provided by the organizer.

## 3. Experiments

### 3.1 Experimental setup

In the FSR-2020 challenge, the training dataset released by the organizer for developing our ASR systems is TAT-Vol1 [15], which is a publicly-available Taiwanese speech corpus. We made use of the whole TAT-Vol1 corpus as our training dataset (about 41 hours) and evaluated our system and various modeling approaches with the pilot-test dataset (about 5 hours; released by the organizer for the warm-up evaluation). To further utilize off-the-shelf audio data, we adopted NER corpus [26] into this work, which contains about 360 hours Taiwanese Mandarin speech. The detailed statistics of the corpus are summarized in Table 1, and all the ASR systems with different modeling approaches were developed with the Kaldi toolkit [16]

| TAT-Vol1 (Taiwanese) | | | |
|---|---|---|---|
| | Speakers | Utterances | Duration |
| Train | 80 | 22,605 | 41 hours |
| Pilot-test | - | 2,617 | 5 hours |
| Final-test | - | 5,663 | 10 hours |
| **NER (Taiwanese Mandarin)** | | | |
| | Speakers | Utterances | Duration |
| Vol1-2-3 | - | 57,387 | 360 hours |

Table 1: The statistics of TAT-Vol1 and NER corpus.

In order to verify the effectiveness of our modeling approaches, we evaluated our systems with two distinct metrics for front-end SE and back-end ASR, respectively. In the front-end experiments, we will evaluate our SE component with the scale-invariant signal-to-noise ratio (SI-SNR) metric, which has been previously shown to be closely related to recognition error reduction [17]: the higher the SI-SNR score the better the ASR performance. On the other hand, we will use syllable error rate (SER) followed by the rules suggested by [15] to evaluate our back-end ASR systems with different modeling approaches. All the SER results depicted in the following experiments will take into account the correctness of the tone index attached to each syllable.

**3.2 Experiments on data cleansing and lexicon augmentation**

| Acoustic Model | Parameters | Data Cleansing | Lexicon Augmentation | SER (%) |
|---|---|---|---|---|
| TDNN-F | 17M | - | - | 19.21 |
| TDNN-F | 17M | × | - | 17.16 |
| TDNN-F | 17M | × | × | 17.15 |
| TDNN-F(M) | 19M | × | × | **16.73** |
| TDNN-F(L) | 21M | × | × | 17.24 |

Table 2: SER (%) results on the pilot-test dataset with different baseline acoustic models.

In the first set of experiments, we intend to examine the two preprocessing approaches mentioned in Section 2.1, namely data cleaning and lexicon augmentation, whose corresponding results are shown in Table 2. As can be seen from Table 2, when the data cleansing approach is applied, our DNN-HMM system can yield a relative SER reduction of 10.7% compared to the baseline system (TDNN-F) provided by the organizer of the FSR-2020 challenge. By comparison, when data cleaning is further paired with the lexicon augmentation, only a moderate improvement can be obtained. We hence conjecture that data cleansing is an indispensable component in the preprocessing stage for acoustic modeling. In addition, we also conduct a model ablation study to check whether stacking more layers to form a deeper neural network for acoustic modeling can lead to better performance. We find that it is not always the case when we add more layers to form a deeper neural network for acoustic modeling. On top of the best result drawn from Table 2, we will use TDNN-F(M) as our default acoustic model for the following experiments.

### 3.3 Experiments on data augmentation

| Noise Injection | SpecAugment | Training epochs | SER (%) |
| --- | --- | --- | --- |
| - | - | 6 | 16.73 |
| × | - | 6 | 16.36 |
| × | × | 6 | 15.22 |
| × | × | 12 | **14.68** |

Table 3: SER (%) results on the pilot-test dataset with disparate data augmentation methods.

In addition to the speed perturbation technique that is employed in the baseline setup, we also consider the use of another two data augmentation methods in building our ASR systems, viz. noise injection and spectrogram augmentation (*cf.* Section 2.2). Notably, we used our default acoustic model for this purpose, while the corresponding results are depicted in Table 3. It is evident from these results that the inclusion and combination of these two data augmentation methods can considerably boost the ASR performance. Among other things, due to the underfitting problem incurred by SpecAugment [5], a further increase of the training epochs from 6 to 12 leads to the best relative SER reduction of 12.2% in relation to the baseline system.

## 3.4 Experiments on acoustic and language modeling

| Acoustic Model | SER (%) | |
|---|---|---|
| | pilot-test | noisy-pilot-test |
| Baseline (TDNNF) | 14.68 | 26.49 |
| + Four-gram LM | **13.47** | 25.13 |
| E2E-TDNNF | 16.09 | 24.08 |
| + Four-gram LM | 14.58 | **21.71** |

Table 4: SER (%) results on the pilot-test dataset with different combinations of the acoustic and language models.

To make our pilot-test dataset consistent with the final test dataset which will be corrupted by varying noise sources, we also inject different types of noise, with SNR levels ranging from 5 to 20 dB, into the pilot-test dataset to form a noisy version of the training dataset that can be additionally exploited for training the acoustic model. Furthermore, we also evaluate two kinds of acoustic models, viz. baseline (TDNNF) and E2E-TDNNF (*cf.* Section 2.4), with the same model architecture and the best data augmentation setting made in Table 3. The corresponding results are shown in Table 4, from which an interesting phenomenon can be observed: when with the clean-condition testing setup, E2E-TDNNF performs worse than TDNNF that is based on the regular training configuration. On the contrary, in the noisy-condition testing setup, the E2E-TDNNF demonstrates superior noise-robustness performance.

On the other direction, when we adopt the four-gram language model (in replace of the trigram language model) for the ASR system to decode syllable sequences, a significant SER reduction can be obtained (*cf.* the last two rows of Table 4). As a side note, we also put effort into training variants of the recurrent neural network (RNN)-based language model for second-pass lattice rescoring [12][13][14]. Their performance, however, was not as good as expected, and we thus omit the details on the lattice rescoring experiments. From now on, unless otherwise stated, all variants of our ASR system discussed in the following experiments will use the four-gram language model for the decoding of syllable sequences.

## 3.5 Front-end speech enhancement

| SE model | SI-SNR (dB) | |
| --- | --- | --- |
| | dev | noisy-pilot-test |
| No processing | 6.82 | 7.26 |
| Conv-TasNet [6] | **16.43** | **17.37** |

Table 5: SI-SNR (dB) results on the noisy pilot-test dataset with the front-end SE component.

| Acoustic Model | SER (%) | |
| --- | --- | --- |
| | noisy-pilot-test | |
| | no process | enhanced |
| E2E-TDNNF | 21.71 | 20.93 |
| E2E-TDNNF-ENH | **20.76** | **19.57** |

Table 6: SER (%) results on the noisy pilot-test dataset with the front-end SE component.

In an attempt to confirm the noise-robustness ability of our ASR system, we conduct a set of experiments with the front-end SE method, viz. Conv-TasNet, which aims at noise suppression. To train Conv-TasNet, we randomly set aside a portion of the training dataset as the development set, and followed the best criterion for training Conv-TasNet that was suggested by [6], viz. minimization of the negative SI-SNR loss. Note here that the so-called permutation invariant training (PIT) was not employed. As can be seen from Table 5, when Conv-TasNet is applied, the SI-SNR results on both the development and pilot-test datasets can be considerably improved. Meanwhile, the SER performance of our ASR system can be substantially promoted (*cf.* the first row of Table 6).

To further enhance the acoustic modeling of our ASR system, we additionally augment the training dataset with a copy of the noisy training utterances which was processed by Conv-TasNet. As such, the augmented training dataset includes the original training data released by the organizer, its noise-contaminated counterpart and its noise-contaminated counterpart further enhanced by Conv-TasNet. We refer to this acoustic model as "E2E-TDNNF-ENH" in contrast to the original one (viz. E2E-TDNNF); E2E-TDNNF-ENH is created to simulate the effect of retraining of the acoustic model on enhanced speech signals. Inspection of the last row of Table 6, we can see that E2E-TDNNF-ENH can bring about a substantial SER improvement in comparison to E2E-TDNNF when the test utterances to be fed into the acoustic model were also enhanced by Conv-TasNet a priori. Nevertheless, in our experiments we spot-checked a few utterances of the pilot-test dataset that were enhanced by Conv-TasNet, and found that Conv-TasNet sporadically eliminated the speech portions of test utterances (viz. most of the speech portions became silent), which probably would lead to deteriorated ASR performance on the unseen test utterances. To secure a reliable performance level of our ASR system on the final test dataset, Conv-TasNet was merely used to obtain an enhanced-copy of the noisy training data, with the purpose of data augmentation for training E2E-TDNNF-ENH. Namely, Conv-TasNet will not be used to enhance the utterances of the final test dataset.

### 3.6 Transfer learning and semi-supervised learning

| Acoustic Model | SER (%) |
|---|---|
| | noisy-pilot-test |
| TDNNF-NER | 25.15 |
| CNN-TDNNF-NER | 23.69 |
| CNN-TDNNF-NER (with semi-supervised training) | **22.68** |

Table 7: SER (%) results on the pilot-test dataset with transfer learning and semi-supervised learning.

In this paper, we also seek to capitalize on more training techniques for acoustic modeling in the context of Taiwanese ASR. To this end, we adopt the strategy proposed in [23], which in essence involved two techniques: transfer learning [24] and semi-supervised training [25]. In implementation, we first used the weight transfer strategy [24] to train an acoustic model with parts of its model parameters transferred from a source model that were well-trained on the NER dataset [26] beforehand. On a separate front, we also attempt to make use the label-agnostic final test dataset (viz. the corresponding reference transcripts of the final test dataset were not provided) to perform semi-supervised training of the acoustic model. In implementation, the recipe proposed in [25] was adopted, which used the entire lattice pertaining to each unlabeled training utterance as the supervision. The corresponding results are shown in Table 7, from which several observations can be drawn. First, when the TDNNF-based acoustic model was trained with transfer learning (denoted by TDNNF-NER), the SER result is slightly degraded compared to the result (25.13%) listed in the second row of Table 4. Second, if the acoustic model was built on top of the CNN-TDNNF structure, transfer learning can offer a considerable SER improvement on the noisy pilot-test dataset (*cf.* the second row of Table 7). It should be noted here that the CNN-TDNNF-based acoustic model, merely trained on the 50-hour training dataset offered by the FSR-2020 challenge, yields SER results significantly lower than 25.13%. This to some extent reveal that as opposed to TDNNF, CNN-TDNNF requires a larger amount of training dataset to achieve better ASR performance. In addition, when the label-agnostic final test dataset was additionally exploited to fine-tune the acoustic model, the performance of our ASR system on the pilot-test dataset can be boosted by a significant margin.

## 3.7 System combination

| Combined systems | SER (%) | |
| --- | --- | --- |
| | noisy-pilot-test | final-test |
| + CNN-TDNNF-NER<br>+ E2E-TDNNF<br>+ E2E-TDNNF-ENH | 19.33 | 13.60 |
| + CNN-TDNNF-NER (Semi-supervised)<br>+ E2E-TDNNF<br>+ E2E-TDNNF-ENH | **19.10** | **13.10** |

Table 8: SER (%) results achieved by our two system-ensemble approaches on the noisy pilot-test dataset and final-test dataset.

In the last set of experiments, we report on the results of our ASR systems submitted to FSR-2020 challenge, which were built based on two system-ensemble approaches that make combinations of different ASR systems previously evaluated in the above subsections. To be specific, we first performed lattice combination to merge all of the word lattices generated by different ASR systems into a single one with equal prior weights. Then, minimum Bayes-risk (MBR) decoding was conducted to obtain the ultimate ASR output for each test utterance. Here we combine the first three of the best systems according to their performance on the noisy pilot-test dataset. Table 8 shows the SER results of our two system-ensemble approaches on the noisy pilot-test dataset and final-test dataset. It is clear that these two system-ensemble approaches can substantially improve the ASR performance of our system on the pilot-test dataset.

## 3.8 Summary of the experiments

Finally, we summarize the SER results of the participating teams on the final test dataset of Track 3 in the FSR-2020 challenge. Figures 2 and 3 show the SER evaluations of Track 3 with and without consideration of the correctness of tone transcription, respectively. Note here that although each team could submit two disparate results for evaluations, we only list the best result of each team here for brevity. Our ASR system has achieved the best performance among all participating teams for the two evaluation settings.

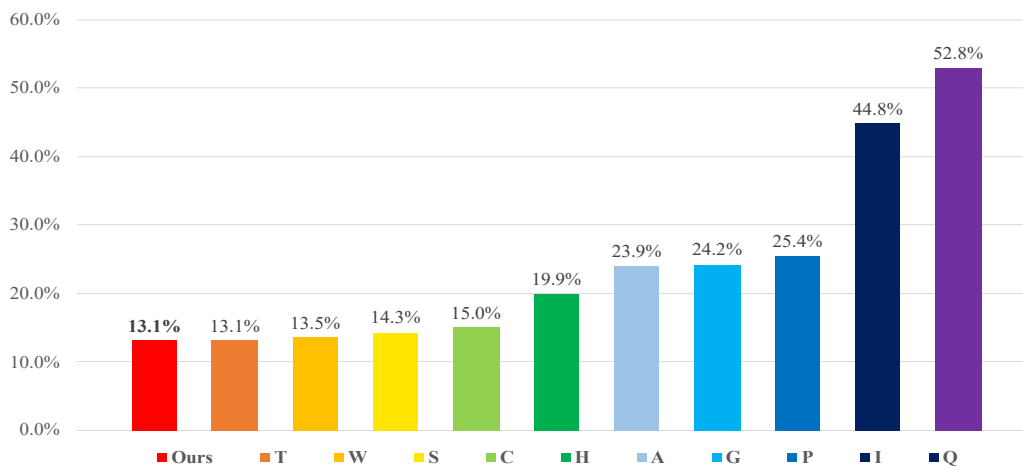

Figure 2: SER (%) results of all participating teams submitted to Track 3 in the FSR-2020 challenge (with consideration of the correctness tone transcription).

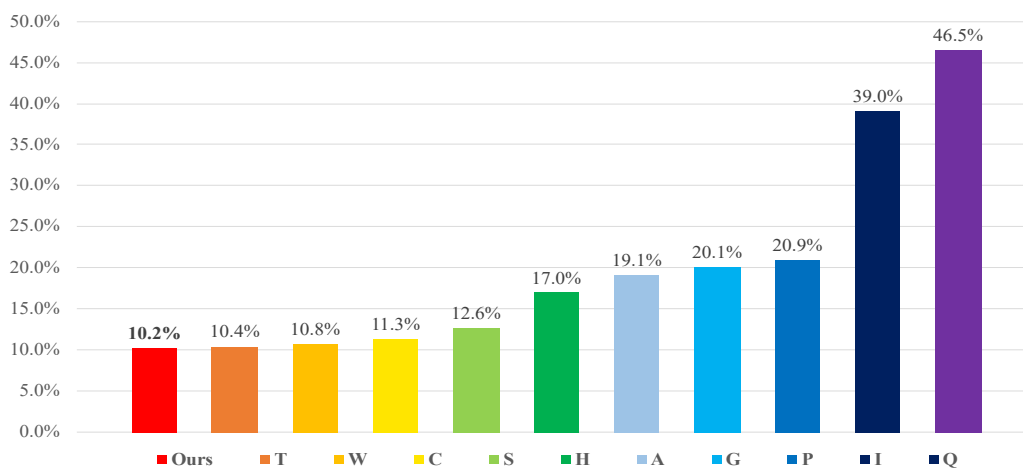

Figure 3: SER (%) results of all participating teams submitted to Track 3 in the FSR-2020 challenge (without consideration of the correctness tone transcription).

# 4. Conclusion and Future Work

In this paper, we have presented the modeling details of the NTNU ASR system that participated in the FSR-2020 Challenge. Through a series of experimental evaluation, the promising effectiveness of the joint use of data cleansing, data augmentation, front-end approach, transfer learning and semi-supervised learning methods for Taiwanese speech ASR has been confirmed. As to the future work, we plan to investigate more sophisticated end-to-end approaches for use in acoustic modeling of the Taiwanese language, as well as to apply our modeling strategy to other resource-poor ASR tasks.

# References

[1] "Taiwanese Hokkien - Wikipedia." [Online]. Available:

*https://en.wikipedia.org/wiki/Taiwanese_Hokkien*

[2] D. Povey, G. Cheng, Y. Wang, K. Li, H. Xu, M. Yarmohammadi, and S. Khudanpur, "Semi-orthogonal low-rank matrix factorization for deep neural networks," in *Proc. Interspeech*, pp. 3743–3747, 2018.

[3] T. Ko, V. Peddinti, D. Povey, and S. Khudanpur, "Audio augmentation for speech recognition," in *Proc. Interspeech*, pp. 3586-3589, 2015.

[4] N. Jaitly and G. E Hinton, "Vocal tract length perturbation (VTLP) improves speech recognition," in *Proc. ICML*, 2013.

[5] D. S. Park, W. Chan, Y. Zhang, C. -C. Chiu, B. Zoph, E. D. Cubuk, and Q. V. Le, "SpecAugment: A simple data augmentation method for automatic speech recognition," in *Proc. Interspeech*, pp. 2613- 2617, 2019.

[6] Y. Luo and N. Mesgarani, "Conv-TasNet: Surpassing ideal time–frequency magnitude masking for speech separation," *IEEE/ACM Transactions on Audio, Speech, and Language Processing*, vol. 27, no. 8, pp. 1256–1266, 2019.

[7] D. Povey, V. Peddinti, D. Galvez, P. Ghahremani, V. Manohar, X. Na, Y. Wang, and S. Khudanpur, "Purely sequence-trained neural networks for ASR based on lattice-free mmi," in *Proc. Interspeech*, pp. 2751–2755, 2016.

[8] H. Hadian, H. Sameti, D. Povey, and S. Khudanpur, "End-to-end speech recognition using lattice-free MMI," in *Proc. Interspeech*, 2018.

[9] A. Stolcke, "SRILM - an extensible language modeling toolkit," in *Proc. ICASSP*, 2002.

[10] W. A. Gale, "Good-Turing smoothing without tears," *Journal of Quantitative Linguistics*, 1995.

[11] H. Ney, U. Essen, "On smoothing techniques for bigram-based natural language modelling," In *Proc. ICASSP*, pp. 825–828, 1991.

[12] H. Xu, K. Li, Y. Wang, J. Wang, S. Kang, X. Chen, D. Povey, and S. Khudanpur, "Neural network language modeling with letter-based features and importance sampling," in *Proc. ICASSP*, pp. 6109–6113, 2018.

[13] C. Wang, M. Li, and A. J. Smola, "Language models with transformers," *arXiv preprint arXiv:1904.09408*, 2019.

[14] S. -H. Chiu and B. Chen, "Innovative BERT-based reranking language models for speech recognition," in *Proc. SLT*, 2021.

[15] Y. -F. Liao *et al.*, "Formosa Speech Recognition Challenge 2020 and Taiwanese Across Taiwan Corpus," in *Proc. O-COCOSDA*, 2020.

[16] D. Povey, A. Ghoshal, G. Boulianne, L. Burget, O. Glembek, N. Goel, M. Hannemann, P. Motlicek, Y. Qian, P. Schwarz, J. Silovsky, G. Stemmer, K. Vesely, "The Kaldi speech recognition toolkit," in *Proc. ASRU*, 2011.

[17] K. Kinoshita, T. Ochiai, M. Delcroix, and T. Nakatani, "Improving noise robust automatic speech recognition with single-channel time-domain enhancement network," in *Proc. ICASSP,* pp. 7009–7013, 2020.

[18] D. Snyder, G. Chen, and D. Povey, "MUSAN: A music, speech, and noise corpus," *arXiv preprint arXiv:1510.08484*, 2015.

[19] J. Thiemann, N. Ito, and E. Vincent, "The diverse environments multichannel acoustic noise database: A database of multichannel environmental noise recordings," *The Journal of the Acoustical Society of America*, vol. 133, no. 5, pp. 3591–3591, 2013.

[20] D. B. Dean, S. Sridharan, R. J. Vogt, and M. W. Mason, "The QUT-NOISE-TIMIT corpus for the evaluation of voice activity detection algorithms," in *Proc. Interspeech*, pp. 3110–3113, 2010.

[21] F. Saki, A. Sehgal, I. Panahi, and N. Kehtarnavaz, "Smart phone-based real-time classification of noise signals using subband features and random forest classifier," in *Proc. ICASSP*, pp. 2204–2208, 2016.

[22] F. Saki and N. Kehtarnavaz, "Automatic switching between noise classification and speech enhancement for hearing aid devices," in *Proc. EMBC*, pp. 736–739, 2016.

[23] T. -H. Lo and B. Chen, "Semi-supervised training of acoustic models leveraging knowledge transferred from out-of-domain data," in *Proc. APSIPA ASC*, pp. 1400-1404, 2019.

[24] P. Ghahremani, V. Manohar, H. Hadian, D. Povey, and S. Khudanpur, "Investigation of transfer learning for ASR using LF-MMI trained neural networks," in *Proc. ASRU*, pp. 279–286, 2017.

[25] V. Manohar, H. Hadian, D. Povey, S. Khudanpur, "Semi-supervised training of acoustic models using lattice-free MMI," in *Proc. ICASSP*, 2018.

[26] Y. -F. Liao, Y. -H. Shawn Chang, S. -Y. Wang, J. -W. Chen, S. -M. Wang and J. -H Wang, "A progress report of the Taiwan Mandarin Radio Speech Corpus Project," in *Proc. O-COCOSDA*, 2017.